\documentclass[10pt, conference]{IEEEtran}
\usepackage[noend]{algpseudocode}
\usepackage[linesnumbered,ruled,vlined]{algorithm2e}
\usepackage{booktabs}
\usepackage{cite}
\usepackage{comment}
\usepackage{amsmath,amssymb,amsfonts,amsthm}
\usepackage{csquotes}
\usepackage{graphicx}
\usepackage[backref]{hyperref}
\usepackage{cleveref}
\usepackage[autolanguage]{numprint}
\usepackage{textcomp}
\usepackage{tikz}
\usetikzlibrary{patterns}
\usetikzlibrary{positioning}
\usepackage{pdflscape}
\usepackage{subcaption}
\usepackage{xcolor}
\usepackage{xspace}
\usepackage{balance}
\usepackage{todonotes}
\usepackage{soul}
\usepackage{multirow}
\usepackage{etoolbox}

\newcommand{\ie}{i.\,e.,\xspace}
\newcommand{\eg}{e.\,g.,\xspace}

\newcommand{\etal}{et al.\xspace}
\newcommand{\quot}[1]{``#1''}

\newcommand{\Oh}{\ensuremath{\mathcal{O}}}

\newcommand{\tool}[1]{{\small \textsf{#1}}}
\newcommand{\ext}{\tool{\geo-R}}
\newcommand{\geo}{\tool{Geographer}}
\newcommand{\geoKM}{\tool{geoKM}}

\newcommand{\geoRef}{\tool{geoRef}}
\newcommand{\geoPM}{\tool{geoPMRef}}
\newcommand{\parm}{\tool{ParMetis}}
\newcommand{\parmgr}{\tool{ParMetisGraph}}
\newcommand{\parmge}{\tool{ParMetisGeom}}
\newcommand{\parh}{\tool{ParHip}}
\newcommand{\xpulp}{\tool{xtraPulp}}
\newcommand{\mj}{\tool{MultiJagged}}
\newcommand{\km}{$k$-means}

\newcommand{\ptsc}{\tool{PT-Scotch}}

\newcommand{\metis}{\tool{Metis}}
\newcommand{\kh}{\tool{KaHip}}
\newcommand{\scot}{\tool{Scotch}}

\newcommand{\topoA}{{\sc topo1}\xspace}
\newcommand{\topoB}{{\sc topo2}\xspace}
\newcommand{\topoC}{{\sc topo3}\xspace}

\newcommand{\cs}{c_{s}}
\newcommand{\Cs}{C_{s}}
\newcommand{\mcap}{m_{cap}}
\newcommand{\Mcap}{M_{cap}}
\newcommand{\creq}{c_{req}}
\newcommand{\Creq}{C_{req}}
\newcommand{\js}{\texttt{jSpeed}}
\newcommand{\cl}{\texttt{jLoad}}
\newcommand{\tw}{tw}
\newcommand{\desW}{\texttt{desW}}

\newtheorem{theorem}{Theorem}

\newtheorem{lemma}{Lemma}

\Crefname{Figure}{Fig.}{Figs.}
\crefname{figure}{fig.}{figs.}

\definecolor{thirdhue}{RGB}{31,120,180}

\newcommand{\hmey}[1]{{\color{orange}HM: #1}}

\newcommand{\ct}[1]{{\color{thirdhue}{[HT: #1]}\xspace}}

\newcommand{\done}[1]{}

\Crefname{equation}{Eq.}{Eqs.}	

\graphicspath{{./figures/}}

\newtoggle{arxiv_ver}
\toggletrue{arxiv_ver}		

\begin{document}

\title{
Distributing Sparse Matrix/Graph Applications in Heterogeneous Clusters -- an Experimental Study
}


\author{
\IEEEauthorblockN{Charilaos Tzovas}
\IEEEauthorblockA{\textit{Humboldt-Universit\"at zu Berlin}\\
\textit{Department of Computer Science} \\
Berlin, Germany \\
tzovas.charilaos@hu-berlin.de}
\and
\IEEEauthorblockN{Maria Predari}
\IEEEauthorblockA{\textit{Humboldt-Universit\"at zu Berlin}\\
\textit{Department of Computer Science} \\
Berlin, Germany \\
predarim@hu-berlin.de}
\and
\IEEEauthorblockN{Henning Meyerhenke}
\IEEEauthorblockA{\textit{Humboldt-Universit\"at zu Berlin}\\
\textit{Department of Computer Science} \\
Berlin, Germany \\
meyerhenke@hu-berlin.de}
}

\maketitle
\IEEEpeerreviewmaketitle

\pagestyle{plain}

\begin{abstract}
Many problems in scientific and engineering applications contain sparse matrices or graphs
as main input objects, \eg numerical simulations on meshes. Large inputs are abundant these
days and require parallel processing for memory size and speed. 
To optimize the execution of such simulations on cluster systems, the input problem needs 
to be distributed suitably onto the processing units (PUs). 
More and more frequently, such clusters contain different CPUs or a combination of CPUs and GPUs.
This heterogeneity makes the load distribution problem quite challenging.
Our study is motivated by the observation that established partitioning
tools do not handle such heterogeneous distribution problems as well as homogeneous ones.

In this paper, we first formulate the problem of balanced load distribution for
heterogeneous architectures as a multi-objective, single-constraint
optimization problem. We then split the problem into two phases and propose a greedy approach to
determine optimal block sizes for each PU.
These block sizes are then fed into numerous existing graph partitioners,
for us to examine how well they handle the above problem.
One of the tools we consider is an extension of our own previous work (von Looz \etal, ICPP'18)
called \geo.
Our experiments on well-known benchmark meshes indicate that only two tools
under consideration are able to yield good quality.
These two are \parm\ (both the geometric and the combinatorial variant) and \geo.
While \parm\ is faster, \geo\ yields better quality on average.
\end{abstract}



\section{Introduction}
\label{sec:intro}
%
Applications with sparse matrices or sparse graphs are ubi\-quitous in science and engineering.
In case of sparse matrices, applications often model complex problems as discretizations of 
partial differential equations, \eg in molecular dynamics~\cite{Mniszewski15}
or climate simulations~\cite{Duchaine17}. This usually leads to large matrix-vector
problems such as sparse linear systems or eigenproblems with significant
demands in terms of memory and computation. Hence, such simulations are often
executed on parallel systems with distributed memory. Typical algorithms
include Krylov subspace solvers such as conjugate gradient (CG),
which are iterative and perform sparse matrix-vector product (SpMV)
operations in each iteration.
Due to the correspondence between matrices and graphs~\cite{DBLP:conf/hpec/KepnerABBFGHKLM16},
similar computational demands arise in large-scale graph computations such
as the analysis of massive online social 
networks~\cite{DBLP:journals/pvldb/ChingEKLM15}.\footnote{In the remainder, we focus in the
description of the underlying application on matrix computations, but graph
computations are equally relevant.}

To obtain short execution times, it is essential to find a good distribution
of the application's computational tasks onto the available processing 
units (PUs). This is all the more important for heterogeneous systems, \ie
when the PUs differ in terms of their speed and memory capacity. Such systems
become more and more common since GPUs offer a more energy-efficient way to
solve certain problems in parallel~\cite{zhang2015efficient}.

A good distribution balances the load (in particular the matrix/graph and
the main vectors involved in the computations) between the PUs and leads to a low
communication overhead. Since for many of the problems under consideration,
GPUs are faster than CPUs, they could in principle receive a larger
share of the distributed data. At the same time, however,
GPUs usually have smaller memories -- a constraint that must be obeyed
to guarantee a healthy application.

One popular way of finding a good distribution is to use graphs and/or
hypergraphs to model the task interactions and then to employ (hyper)graph/matrix
partitioning~\cite{DBLP:series/lncs/BulucMSS016}. Finding an \emph{optimal} distribution
this way is $\mathcal{NP}$-hard~\cite{Garey1979}.
Thus, in practice heuristics are applied~\cite{HendricksonLeland95multilevel,KernighanLin70efficient,Bui93}.
(Hyper)Graph/matrix partitioning usually leads to a partition of the graph/matrix
into so-called \emph{blocks} of vertices/rows, where each block is handled by one PU. 
Since classic graph partitioning does not take into account how fast the PUs can communicate with each other, 
explicit algorithms for mapping the blocks onto PUs are sometimes employed 
in addition, see~\cite{bichot2013graph}. This mapping step becomes more important for
more complex/irregular scenarios~\cite{DBLP:conf/icpp/GlantzPM18} 
(\eg in terms of application matrices, heterogeneous and/or hierarchical compute systems).


\paragraph*{Motivation}
From an abstract perspective, the goal is to (re)distribute the input matrix 
among the available PUs in a way such that the application's execution time
is minimized. This means that (i) no PU receives more data than it can fit 
into its memory, (ii) each PU receives a part of the input that matches its
proportionate computational capabilities, and (iii) the communication costs
induced by the distribution are low.
Partitioning tools that offer mapping capabilities are addressing requirement (iii)
and often result in distributions of low communication costs.
However, most of these tools do not provide automatic support for heterogeneous 
PU characteristics, thus, they do not address requirements (i) and (ii).
Indeed, modern systems with CPUs and GPUs
working on the same problem together lead to conflicting demands in terms of 
processing capabilities and available memory.
The lack of explicit support by existing partitioning tools for the
above problem is what motivates our present study.

\paragraph*{Outline and Contribution}

In Sec.~\ref{sec:problem} we formulate the problem of balanced load distribution
for heterogeneous PUs as multi-objective optimization problem with a memory constraint.
To be able to use existing partitioners (which are reviewed in Sec.~\ref{sec:relwork}),
we divide the problem into two stages: first, we compute the desired block sizes with 
a greedy method in $\Oh(k \log k)$ time for $k$ PUs, see Sec.~\ref{sec:blocksizes}.
There, we also prove these block sizes to be optimal for this first stage (but not for
the complete optimization problem, of course).
The block sizes are then fed into a wide variety of popular partitioning tools 
to address the second objective in a second stage.
After excluding three tools for not supporting our problem adequately,
our experimental study on the remaining three tools with eight algorithms
(Sec.~\ref{sec:experiments}) shows: only \parm\ and \geo\ 
handle the heterogeneous load distribution with good solution quality.
A combination of balanced \km\ and combinatorial refinement, introduced here in \Cref{sec:geo-exte},
calculates partitions for meshes with $10\%$ better cut value compared to \parm.
In most cases, we \quot{pay} only with a running time increase of about $50\%$,
and there are several cases where \geo\ is both faster \emph{and} has better quality.
\iftoggle{arxiv_ver}
{\textbf{Material omitted from the main part can be found in
    the appendix.}
}
{
\textbf{Material omitted due to space constraints can be found in
  the appendix of the full version of this paper~\cite{fullVersion}.}
}

\section{Problem Statement}
\label{sec:problem}
As is customary, we exploit the correspondence between a symmetric 
$n \times n$ matrix $A$ and an undirected graph $G$ with $n$ vertices~\cite{DBLP:conf/hpec/KepnerABBFGHKLM16}.
$G$ has an edge $\{u,v\}$ iff the entry $A[u,v]$ is different from zero. 
This allows to see the load distribution problem for matrices through
graph lens. A list of important symbols and acronyms that are used in the
remainder are given in \Cref{app:list-of-symbols}.
\begin{table}[b]
\centering
\begin{small}
\begin{tabular}{cp{0.39\textwidth}}

  Symbol & Description \\
  \hline
  GP & graph partitioning \\
  LDHT & load distribution problem for heterogeneous topologies \\
  $k$ & number of blocks of a partition \\
  $\varepsilon$ & partition imbalance threshold \\
  $cut(\cdot)$ & number of edges cut in a partition \\
  $b_i$ & block $i$ of a partition \\
  $p_i$ & processing unit $i$  of the system (PU) \\
  $\cs(\cdot)$ &  speed of a PU (number of operations per time unit) \\
  $\mcap(\cdot)$ & memory capacity of a PU \\
  $\Cs$ & total computational speed of the system \\
  $\Mcap$ & total memory of the system\\
  $tw(\cdot)$  & the target weight of a block
\end{tabular}
\end{small}
\caption{Description of symbols used in the paper.}
\label{app:list-of-symbols}
\end{table}

\subsection{Graph Partitioning (GP)}
\label{sub:graph-part}
We assume an application where each matrix element / graph vertex represents the same computational demand and memory requirement
(thus both are normalized to $1$).
As a first step, let us also assume a homogeneous compute system
in which all PUs have the same speed. Moreover, let them communicate among each other
with the same speed, regardless of the pair of PUs under consideration.

To make the application as fast as possible, we want to balance the load
among PUs and minimize the (expensive) communication between PUs.
In such a setting, the balanced load distribution problem is usually modeled by
\emph{graph partitioning} (GP). For GP, given an application graph $G_a = (V, E)$, 
an integer $k$ and an imbalance threshold $\varepsilon$, we seek a partition
$\Pi$ into $k$ blocks $b_0,b_1,\dots b_{k-1}$ such that a cost function $f(\Pi)$
is minimized, while for all blocks $|b_i|\leq (1+\varepsilon)n/k$ holds.
In case of unweighted vertices or unit weights, $|b_i|$ denotes the number 
of nodes in block $b_i$.

The most common cost function is the edge cut, $cut(\Pi)$, \ie the number (or weight) 
of edges with endpoints in different blocks, see~\cite{DBLP:series/lncs/BulucMSS016}. Some authors also considered the (more 
accurate) communication volume and/or the number of boundary vertices, \ie 
the number of vertices with at least one neighbor in a different block~\cite{HendrichsonKolda00}.

\if #0
\ct{
\paragraph*{Extensions}

The problem can be extended in many ways. A simple version is to allow vertices to
have a weight $w(v)$. In this case, the size of a block is $|b_i|=\sum_{v\in b_i} w(v)$,
the total weight of the graph is $W$ and one requires $|b_i|\leq (1+\varepsilon)W/k$ for all $i$.

\hmey{I guess we don't do the following in detail and simply say that there
exist more detailed variants.}
An even more expressive model is for vertices to have $r$ weights $w_1,\dots,w_r$ 
and we also have $r$ balance constraints $\varepsilon_1\dots,\varepsilon_r$.
Now, we want all blocks to be balanced for all weights, 
$|b_i^j|=\sum_{v\in b_i} w_j(v)$ for weight $w_j$ and we want
$|b_i^j|\leq (1+\varepsilon_j)W_j/k$ for $j=1,\dots r$.
}
\fi

\subsection{Balanced Load Distribution for Heterogeneous PUs}
\label{sub:het-pus}
%
Next, let us model heterogeneous
PUs. They can have different processing speeds and different memory capacities.
In this scenario, each PU should receive an amount of load that is proportionate
to its speed -- but never more than its memory capacity.
Formally, the input consists of an application graph $G_a = (V, E)$,
the number $k$ of blocks, and a representation of the compute system topology.
Since most compute systems are hierarchical in one way or another,
we make the common assumption~\cite{DBLP:conf/wea/SchulzT17} that this representation takes the form of a tree $T$;
its leaves represent the $k$ PUs, the inner nodes correspond to sets of PUs (\ie to their
descendants in the tree).

Each PU $p_i$ has two weights: $\cs(p_i)$ is its speed (a normalized number
of operations per time unit) and $\mcap(p_i)$ its memory capacity. 
The respective values of an inner node $v$ of $T$ are formed by recursively accumulating
the values of all children of $v$. Thus, $\Cs = \sum_{i=0}^{k-1} \cs(p_i)$ 
and $\Mcap= \sum_{i=0}^{k-1} \mcap(p_i)$ are the total computational speed and memory of the system, respectively.

To be able to assess established tools, we first model the balanced load distribution
problem by computing a partition
$\Pi$ of $V$ into $k$ blocks $b_0,b_1, \dots , b_{k-1}$ again and then mapping
each block $b_i$ to PU $p_i$. 
This means that block $b_i$ must take the speed and the memory
capacity of $p_i$ into consideration.
For the load distribution problem for heterogeneous topologies (LDHT),
the goal is then to find a partition $\Pi$ such that:
\begin{align}
\text{minimize } & cut(\Pi) \text{ and } \label{eq:min-cut} \\
\text{minimize } & \max \frac{tw(b_i)}{\cs(p_i)} \label{eq:min-imb} \\
\text{s.\ t.\ } & tw(b_i) \leq \mcap(p_i), \label{eq:mem-constraint}
\end{align}
where $tw(b_i)$ is the target weight for block $i$.
In this formulation, Eq.~(\ref{eq:min-cut}) is meant to minimize the communication
between PUs, Eq.~(\ref{eq:min-imb}) is meant to balance the computational load
according to the speeds, while the constraint~(\ref{eq:mem-constraint})
enforces that no memory overload occurs.
The LDHT problem is $\mathcal{NP}$-hard since it contains the $\mathcal{NP}$-hard
graph partitioning problem as a special case.




\section{Related Work}
\label{sec:relwork}
In this section, we focus on aspects closely related to the main motivation of our study.
First, we discuss algorithms for the partitioning 
problem, then we consider the related multi-constraint, multi-objective
problem. Finally, we mention existing literature
on load balancing for heterogeneous systems.

\paragraph{GP algorithms}
There is a large amount of literature on heuristics for the GP problem.
Most of them can be classified as either combinatorial, where the connectivity
information of the (hyper)graph is used to drive the partitioning,
or geometric, where solely the coordinate information is used. 
Combinatorial partitioning tools (or simply: (hyper)graph partitioning tools)
usually produce solutions, with better quality
while geometric ones typically have lower running times.
Arguably, one of the most successful heuristics for accelerating
any partitioning algorithm is the multilevel
approach~\cite{HendricksonLeland95multilevel}.
The multilevel approach consists of three phases: coarsening, initial
partitioning, and uncoarsening. 
In the coarsening phase,
vertices are successively merged together to
form a series of smaller graphs 
until a sufficiently small graph is produced.
Then, an initial partition is obtained on that graph 
and is projected back to the original input, through uncoarsening.
In the corresponding uncoarsening level,
the merged vertices are split and the partition
of the coarser graph is further refined.
The multilevel approach is available in a number of well-known graph partitioning tools
such as \metis, \scot~and \kh.
The most commonly used methods for graph coarsening are usually based on vertex-matching
schemes~\cite{karypis1999fast} (as in \metis) and on clusterings from
on label propagation (as used by \xpulp and by some configurations of \kh).
During uncoarsening many tools use local search refinement techniques such FM/KL methods~\cite{Fiduccia82linearMinCut,karypis1999fast}.

Geometric partitioning tools include implementations of space-filling curves~\cite{warren93} (SFC),
recursive inertial bisection~\cite{nouromid1986} (RIB) and
recursive coordinate bisection~\cite{Heath94} (RCB).
RCB is a recursive bisection scheme that attempts to minimize the boundary between the
subdomains by splitting the mesh in half, orthogonal to its longest dimension.
RIB uses the principal inertial axis to
produce a bisection and does not restrict the splitting to one dimension.
Furthermore, a more recent geometric partitioning, called \mj~\cite{DRDC16}, 
computes partitions by recursively  multi-sectioning  the  dataset,
which can also be viewed as a generalization of RCB.
Implementations of all the above methods can be found
in the Zoltan public library~\cite{zoltan}.
We also use \geo, a scalable algorithm for geometric partitioning,
recently proposed by a subset of the authors~\cite{DBLP:conf/icpp/LoozTM18}. \geo\ applies
a balanced version of the $k$-means algorithm to obtain blocks with compact shapes.

Finally, parallel partitioning algorithms are crucial for modern distributed systems,
especially since graph sizes are increasing dramatically.
In general, geometric partitioning tools tend to be easier to parallelize than 
multilevel graph partitioners. However, parallel versions
of all the algorithms/tools presented above are available. Parallel algorithms
exist in \parm, \parh, \xpulp, \ptsc\ for graph partitioning and the
Zoltan library for the geometric ones.
  The classic partitioning algorithms do not explicitly support the LDHT problem.
  However, to solve the LDHT problem, we need a classic partitioning
  tool that accepts specific weights per block. In the experiments, we use 
  the above tools (parallel versions),
  excluding only those who do not allow specific weights as input. 
  Some of these tools are equipped with additional capabilities, such as solutions
  for the multi-constraint, multi-objective problem as \parm~and \xpulp.

\paragraph{Multi-constraint, multi-objective partitioning}
Multi-constraint, multi-objective graph partitioning algorithms
are used to model problems
with several balance constraints
and/or several optimization objectives.
Example problems arise in multi-physics or multi-mesh simulations~\cite{Plimpton98}.
Then, one assigns a weight vector of size $m$ to each vertex and a weight vector
of size $l$ to each edge. The problem then becomes that of finding a partition that
minimizes the edge cut with respect to all $l$ weights,
subject to the constraints that each of the $m$ weights
is balanced across the subdomains.
  \parm\ was one of the first tools to include algorithms
  for the multi-constraint, multi-objective paradigm.
In Ref.~\cite{parmetis} the problem of balancing computation and memory
constraints is modeled as a two-constraint problem 
by associating a vector of size two for each vertex,
where the elements of the vector represent the
computation and memory requirements associated with that vertex.
Later in Ref.~\cite{moulitsas08} the multi-constraint multi-objective algorithm of
\parm\ was used to devise an architecture-aware partitioning algorithm.
  Although solutions for the multi-constraint, multi-objective problem
  are relevant, the existing tools do not explicitly address 
  the LDHT problem; 
   instead, they treat \emph{all} weights as constraints (upper bounds).
  In LDHT,
  the goal is to balance one weight without exceeding the second one.
  We address this problem by describing a two-stage approach. First
  we compute the optimal block weights to be given to these tools, and thus
  our partial problem leads to single-constraint, single-objective graph
  partitioning.

\paragraph{Load balancing for heterogeneous systems}

Related work on load balancing for heterogeneous clusters focuses mainly
on solving the mapping problem (or process/task placement), 
for which the communication costs between PUs are added in
the classic model of graph partitioning.
The majority of those algorithms do not consider heterogeneous clusters
in terms of different speed or memory capabilities (rather only
in terms of communication costs) and thus do not solve the LDHT problem.
More details can be found in a survey~\cite{hoeflerSurvey13}
  on algorithms and software for mapping.
A recent work that focuses on partitioning meshes for heterogeneous systems
appeared in \cite{Chevalier20}.

\section{Determining Optimal Block Sizes}
\label{sec:blocksizes}
Let us assume we are faced with a load distribution problem and use one of
the established tools that accept (nearly)
arbitrary target block sizes/weights $tw(b_0), \dots, tw(b_{k-1})$ for 
the blocks of the application graph's vertex set. Here, we discuss how to 
calculate these $tw(\cdot)$ values for the LDHT problem, which is more difficult than 
standard GP. For GP, one typically passes only $G_a$, $k$, and $\varepsilon$
to the respective tool, which then computes a partition with block sizes
$\leq (1+\varepsilon)\lceil \frac{n}{k} \rceil$. 

For LDHT, the calculation of the $tw(\cdot)$ values is not supported inside 
the respective partitioning tools, though, and needs to happen beforehand.
We thus strive for a two-phase process in which we compute the optimal
block sizes first. These sizes are then passed on to the partitioner.
Of course, a unified process with just one phase would be preferable (regarding quality),
because a two-phase process neglects the objective function edge cut in the first phase
completely. But it would certainly overburden the users of the load distribution tools
to make the necessary deep changes inside the tools (which would need to be
on the algorithmic level). Thus, the cut is optimized by the tools in a second black-box phase
and we focus now on the calculation of the $tw(\cdot)$ values for the LDHT Problem
(unweighted $G_a$, heterogenous topology).

We denote the optimal target block size without any constraints for a block $b_i$ by $\tw(b_i)^*$.
In the trivial case where all PUs have sufficient memory,
this optimal solution (to a much simpler problem) is for every block to 
have a weight proportional to the corresponding PU's speed:
\begin{equation}
\tw(b_i)^* = n \cdot \frac{\cs(p_i)}{\Cs}. 
\label{eq:desWeight}
\end{equation}
%

Of course, in reality each PU has a certain memory capacity that must not be exceeded.
If the ``previously optimal size'' (without constraint) does not fit into the memory of a PU, 
then one should load it as much as possible to obtain the ``new optimum'' (with memory constraint).
Note that a PU with insufficient memory (in comparison to its speed) has an implication
on the load of other PUs as well, since the remaining
total load differs from what we assumed in Eq.~(\ref{eq:desWeight}).

\begin{algorithm}[t]
\KwData{Applic.\ graph $G_a$, $k$, topology info as tree $T$}
\KwResult{Target block sizes $\tw(b_0), \dots, \tw(b_{k-1})$}
Sort PUs such that $\cs(p_0)/\mcap(p_0) \geq \cs(p_1)/\mcap(p_1) \geq \dots \geq \cs(p_{k-1})/\mcap(p_{k-1})$ \label{line:sort-PUs} \\
\cl\ $\gets |V|$ \Comment{computational load of the graph} \label{line:init-compLoad} \\
\js\ $\gets \Cs$ \label{line:after-init} \Comment{normalized speed of all PUs from $T$} \\
\For{$i \gets 0,1,\dots, k-1$}{ \label{line:for-loop}
  \desW$(b_i) \gets \cs(p_i)\cdot \cl\ / \ \js\ $ \label{line:comp-desW} \\
  \Comment{desirable weight proportionate to speed} \\
  \eIf{\desW$(b_i) > \mcap(p_i)$}{
    $tw(b_i) \gets \mcap(p_i)$ \label{line:saturated} \Comment{load = mem capacity} \\
  }
  {
    $tw(b_i) \gets \desW(b_i)$ \label{line:more-mem-than-load}   \Comment{more mem than load} \\
  }
  $\cl\ \gets \cl\ - tw(b_i)$ \label{line:subtract-block-size} \\
  $\js\ \gets \js - \cs(p_i)$ \label{line:adjust-cs} 
}
\Return{$tw(b_0), tw(b_1), \dots, tw(b_{k-1})$}
\caption{Calculate target block sizes for the LDHT Problem.}
\label{alg:homog_input}
\end{algorithm}

For the calculation of the $\tw(\cdot)$, we propose a greedy method,
see Algorithm~\ref{alg:homog_input}. We sort the PUs in decreasing order by the 
criterion $\cs(p_i)/\mcap(p_i)$ (Line~\ref{line:sort-PUs}).
This way, we want to ensure that the fastest PUs get as much load as they can handle.
Lines~\ref{line:init-compLoad} and~\ref{line:after-init} initialize the joint computational
load posed by $G_a$ and the normalized joint speed of the PUs encoded in the topology
tree $T$, respectively. The main loop iterates over the PUs in sorted order.
In Line~\ref{line:comp-desW}, in analogy with Eq.~(\ref{eq:desWeight}), we compute the desirable target weight, $\desW(b_i)$,
of block $b_i$ for PU $p_i$. If $\desW(b_i)$ is larger than the memory capacity
of PU $p_i$, then $p_i$ is assigned this memory capacity and is called \emph{saturated}
(Line~\ref{line:saturated}).
Otherwise, $p_i$'s memory capacity exceeds its desirable target load, so that $p_i$ is assigned exactly
this target load, and is called \emph{non-saturated} (Line~\ref{line:more-mem-than-load}).
Finally, in Lines~\ref{line:subtract-block-size} and~\ref{line:adjust-cs}, we subtract the 
load assigned in this iteration from \cl\ and the current PU's speed from \js, respectively.

Proceeding greedily in this sorted order ensures that we first
fill the ``fast enough'' PUs and they receive as much as they can handle.
Moreover, we prove in the following that one actually obtains the optimal solution of
Eq.~(\ref{eq:min-imb}) under the memory constraints. The algorithm's running time of $O(k \log k)$
is no contradiction to the $\mathcal{NP}$-hardness of LDHT -- due to the separation into two
disjoint phases, we do not solve the LDHT problem in an optimal way.

It is absolutely natural to assume that non-saturated PUs exist -- otherwise no 
valid solution exists. For the optimality proof, we first show that all saturated PUs appear consecutively at
the top of the sorted sequence produced in Line~\ref{line:sort-PUs} 
(the proof can be found in 
\iftoggle{arxiv_ver}{
Appendix~\ref{app:proof-lemma-only-once}).
}
{
Appendix of the full version of the paper\cite{fullVersion}.
}
\begin{lemma}
\label{lem:change-only-once}
No saturated PU appears \emph{after} a non-saturated PU in the sorted sequence produced 
in Line~\ref{line:sort-PUs}.
\end{lemma}

\begin{theorem}
\label{thm:optimal-load}
Algorithm~\ref{alg:homog_input} computes the optimal solution for the 
objective function~(\ref{eq:min-imb}) under the memory constraint~(\ref{eq:mem-constraint}).
It runs in $\Oh(k \log k)$ time.
\end{theorem}

In the inductive optimality proof
\iftoggle{arxiv_ver}{
(Appendix~\ref{app:proof-opt-thm}) 
}
{
(see \cite[Appendix]{fullVersion})
}
we exploit the fact that a change from 
a saturated PU to a non-saturated one in the course of Algorithm~\ref{alg:homog_input}
happens only once -- and one never changes back. This is a consequence from Lemma~\ref{lem:change-only-once}.

\if #0
\paragraph*{Heterogeneous system -- computational weights only}
We look at a more complicated version of the problem where,
the memory requirement for every graph vertex is the same, 
but vertices have computational requirements $\creq(v)$ and $\Creq$ is the total
computational requirement of the input. 
As before, PUs have a memory capacity $\mcap$ and a computational  speed $\cs$
that correlates with the computational requirement of the vertices. 
The whole system has $\Mcap$ memory and $\Cs$ computational speed.

In this version we cannot calculate beforehand the best feasible weights for
each block.
For example, consider that the optimum for a PU $p_i$ is to be assigned a block $b_i$
with $\cs(b_i)=1000$ and has a capacity for $100$ vertices, $\mcap(b_i)=100$.
This can be achieved by 100 vertices, that will have a computational weight of $10$
on average, of by 200 vertices, that will have an average computational weight of $5$. 
We cannot know if such a feasible assignment exists before processing the graph.
\fi

\section{Extensions to \geo}
\label{sec:geo-exte}
In this section we describe a new version of \geo, called \ext, in which we
combine the geometric partitioning with combinatorial techniques.
More precisely, we employ a multilevel approach and use a parallel
variant of the FM algorithm for local refinement.
%
%
Contrary to the classic multilevel approach, we obtain a partition
before even starting to coarsen the graph. This is done to obtain a good initial
data distribution to PUs and we use \geo\ to do so. Each PU receives one block for the
remainder of the procedure, and is responsible for coarsening
its local subgraph using heavy-edge matching techniques.

For the uncoarsening, we create the communication graph $G_c$ (a.\,k.\,a.\ quotient graph:
each vertex of $G_c$ corresponds to a block of $G_a$, weighted edges in $G_c$ model
the communication volume exchanged between the blocks of $G_a$) first.
Inspired by Ref.~\cite{DBLP:conf/ipps/HoltgreweSS10}, 
we compute a maximum edge coloring algorithm to determine the 
communication rounds and the communicating PU pairs in each round.
As a reminder,
the classic FM algorithm sorts vertices by gain and moves the 
vertex with the highest gain if it does not violate the balance constraint.
In our case, we do not consider all vertices but only a smaller number
of those in each block, located in the extended neighborhood of the boundary nodes.
Those vertices are identified by performing a number of BFS rounds
starting from the boundary nodes.
Then, after exchanging candidate nodes among communicating PUs, in each round
we apply pairwise FM refinements between the corresponding
PU pairs in parallel. Each PU in the pair performs the local refinement independently
and the best of the two solutions is kept.
Uncoarsening and local refinement are repeated 
until we obtain a solution for the original graph.

We also extended the balanced \km\ algorithm~\cite{DBLP:conf/icpp/LoozTM18}
from \geo\ to better address hierarchical compute systems.
The hierarchy is provided as an implicit tree by a list of numbers $k_1, k_2, \dots k_h$ for 
$h$ hierarchy levels -- where $k_i$ denotes the tree fan-out on that level.
Thus, the total number of blocks is $k=\prod k_i$.
To account for this hierarchy, we partition on level $i$ each block into
$k_{i+1}$ blocks, respectively. Proceeding in this hierarchical fashion usually
leads to solutions with slightly larger edge cut -- direct $k$-way approaches
often yield better quality than their recursive counterparts in this respect.
The benefit is appreciated, however, from a mapping quality perspective:
blocks that share a border (and thus communicate across this border) will likely be mapped
to nearby PUs. As a fast post-processing step, we do a global repartitioning step that
\quot{smooths} the border and improves the cut.

\begin{figure}
\centering
\includegraphics[scale=0.34]{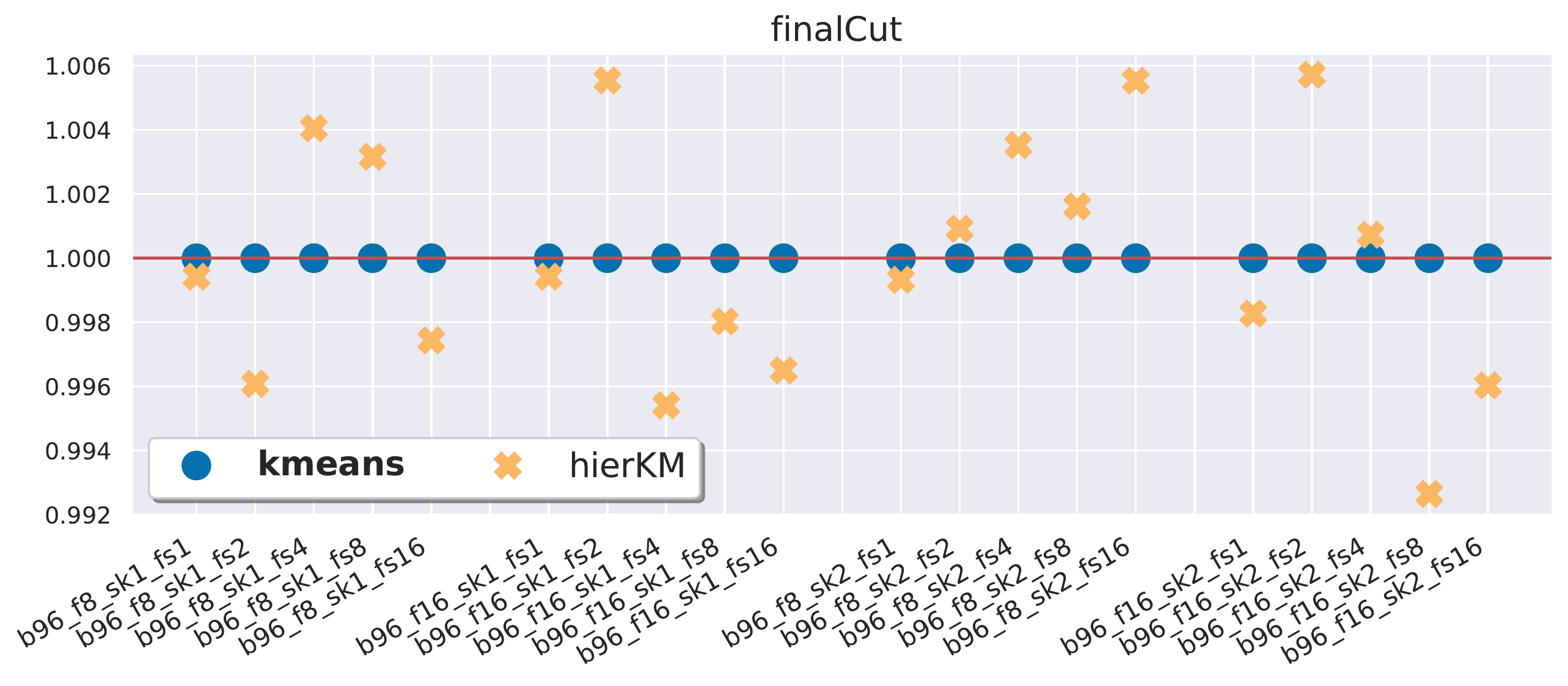}
 
\includegraphics[scale=0.34]{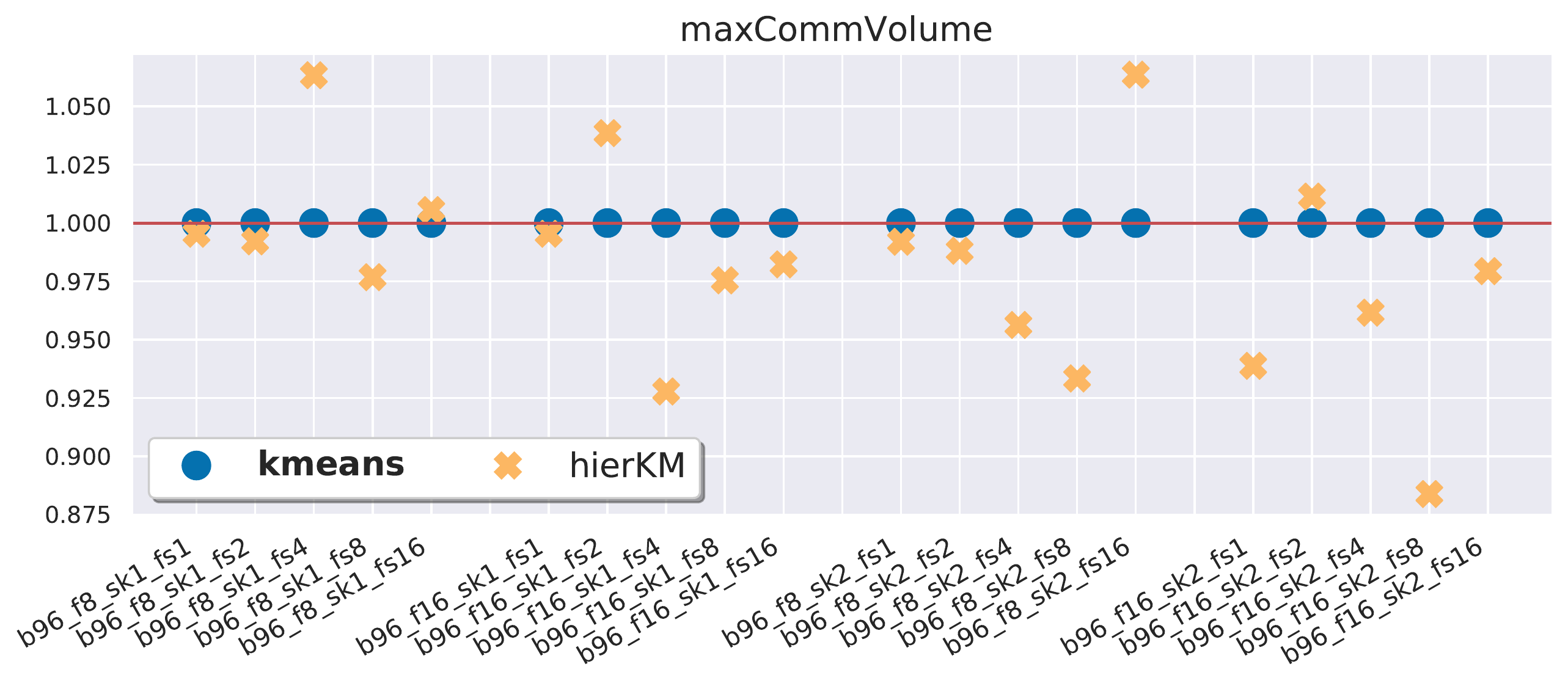}
\caption{Comparison regarding relative quality (top: edge cut, bottom: max.\ communication volume)
between balanced \km\ and the hierarchical version (lower is better).
}
\label{fig:hierGeo}
\end{figure}

Our experiments for the scenario that we emphasize on in this paper indicate that the quality
in terms of edge cut is very close, usually within $\pm 1\%$, see \Cref{fig:hierGeo}.
Although the communication volume can be reduced more in numerous cases,
we decided to provide only the results for the original algorithm in detail to avoid
cluttering the plots.

\section{Experimental Evaluation}
\label{sec:experiments}
In this section, we describe our experiments to evaluate
the effectiveness of different partitioners under
the load distribution model of Sections~\ref{sec:problem} and~\ref{sec:blocksizes},
for various (simulated) heterogeneous systems.
Regarding the requirements of the application,
we only treat the LDHT scenario, where the computational and memory
requirements of the application are equal, corresponding to two unit weights for 
all vertices of the application graph.
For the purpose of our experiments, we simulate
modern systems with different levels of heterogeneity.
To this end, we consider three different topologies; \topoA (see~\ref{exp-topoA}),
\topoB (see~\ref{exp-topoB}) and \topoC (see~\ref{exp-topoC}), which are further
explained in the corresponding subsections.

All relevant material, tool versions and guidelines on how to reproduce the 
experimental results are publicly available in
\url{https://github.com/hu-macsy/geographer/tree/Dev/publication_results}.

\paragraph{Metrics}
Combinatorial quality metrics considered here
are the edge cut and the maximum communication volume.
Regarding actual application performance, we conduct simulations
on a cluster and report running time results for common HPC kernels such as SpMV 
(sparse matrix-vector produce) and linear system solves with CG (conjugate gradient).
We apply the CG solver from the numerical library LAMA~\cite{brandes2017lama} 
to systems derived from the graph's Laplacian matrix.
To ensure that the linear systems have a solution, we shift the diagonal of the Laplacian slightly
to make the matrix positive definite.
For SpMV and for the CG solver, the Laplacian of the input matrix
is distributed according to the partition provided by the respective partitioning tool.

\paragraph{Tool Selection}
For all our experiments we consider a variety of distributed tools for the partitioning
process. More precisely, in the competitors set we include two versions of \parm, 
termed \parmge~and \parmgr. Their difference is that \parmge~uses an SFC to
obtain the initial partition of the graph.
From the tools in the zoltan2 package~\cite{zoltan},
we consider the geometric partitioning methods RCB, RIB, and SFC.
The graph partitioner \xpulp~is not included in our set because it targets complex 
networks and preliminary tests showed insufficient quality (high cut values and 
unbalanced parts) for out data sets.
The current implementation of \mj\ and \parh\ do not accept sufficiently imbalanced block weights,
so that they are excluded as well.

Regarding \geo, \geoKM~refers to the balanced \km~method, 
\geoRef~when we additionally perform local refinement as a postprocessing step.
The version \geoPM~consists of the balanced \km~method plus the local refinement routine
from \parm.

\paragraph{Instances}
To test the partitioners, we use geometric graphs that stem from or resemble
scientific simulations. The graphs are displayed in Table \ref{table:inst}.
The larger graphs were generated
using the generator for random geometric graphs (rgg) and Delaunay triangulations (rdg)
from KaGen~\cite{funke2017}. We also include two graphs from the PRACE benchmark 
suite~\cite{UEABS} that represent the respiratory system.
Moreover, we use the generator from Ref.~\cite{MarquardtSchamberger05open} to create a large
adaptive triangular 2D mesh (from the \emph{refinetrace} series) with $578$M vertices.
The rest of the graphs come from the 10th DIMACS challenge~\cite{BaderMSW12b-dimacs}.
In our detailed presentation of the results, we use aggregate values for representative
instances. Due to the different nature of the individual instances, an aggregation over all
instances would not lead to meaningful results.

\paragraph{Test Systems}
We use two test systems for our experiments. The small one is a local cluster with 16 compute nodes, 
each with 4 6-core Intel Xeon X7460 CPUs and 128 GB of RAM. This system was used for the smaller inputs
from DIMACS and PRACE.
The main volume of the experiments were carried out on 
the HLRN IV system Lise (\url{https://www.hlrn.de/}).
Each compute node of Lise has two Intel Cascade Lake Platinum 9242 CPUs with 384 GB RAM 
and 96 cores.
%
Using two different systems here does not affect the evaluation, since the running time comparison
depends on \emph{relative} values. 
For exact values on certain graph and topologies see \Cref{table:exact_values}.
For the experiments in \topoC, we used the local cluster and tuned down the CPU speed
of certain compute nodes to simulate heterogeneity -- as described in \Cref{exp-topoC}.
In our experiments we assign one MPI process per PU. The factors between different
PU speeds reflect recent results~\cite{chen18} on CPU-GPU comparisons.

\begin{table}
\centering
\begin{tabular}{c|c|c}
name & nodes $n$ & edges \\
\hline
NACA0015 & $\numprint{1039183}$ & $\numprint{3114818}$ \\
M6 & $\numprint{3501776}$ & $\numprint{10501936}$ \\
333SP & $\numprint{3712815}$ & $\numprint{11108633}$ \\
AS365 & $\numprint{3799275}$ & $\numprint{11368076}$ \\
NLR & $\numprint{4163763}$ & $\numprint{12487976}$ \\
hugetric-00020 & $\numprint{7122792}$ & $\numprint{10680777}$ \\
hugetrace-00020 & $\numprint{16002413}$ & $\numprint{23998813}$ \\
hugebubbles-00020 & $\numprint{21198119}$ & $\numprint{31790179}$ \\
alyaTestCaseA & $\numprint{9938375}$ & $\numprint{39338978}$ \\
alyaTestCaseB & $\numprint{30959144}$ & $\numprint{122951408}$ \\
refinetrace & $\numprint{578551252}$ & $\numprint{867786528}$ \\
rdg\_2d\_$2^x$ & $2^x,\; x=25,\dots,29$ & $\approx 3n$ \\
rgg\_2d\_$2^x$ & $2^x,\; x=25,\dots,29$ & $\approx 3n$ \\
rgg\_3d\_$2^x$ & $2^x,\; x=25,\dots,29$ & $\approx 3n$ \\
\end{tabular}
\caption{ The graphs used in the experiments with the number of nodes and edges.}
\label{table:inst}
\end{table}

\subsection{Experiments for \topoA}
\label{exp-topoA}
%
%
In the first category, we have 2 sets of PUs, set $S$ (for slow) and set $F$ (for fast).
This category is further divided into two parts.
Using $k$ as the total number of PUs, we set $|F| := k/12$ in the first part and
$|F| := k/6$ in the second one. The specs of the PUs in $S$ remain always constant
and we increase only the memory capacity and computational speed of the PUs in $|F|$
as follows. 
We start with the same computational speed and memory as the slow PUs;
this corresponds to a homogeneous system since all target block weights are equal.
Then, we increase the computational speed of the fast PUs by a factor of $2$ and
the memory by a factor of $1.6$ for $4$ steps. 
The speed and memory of the PUs in $F$, for each experiment, are shown in 
\Cref{table:topo}.
In the first heterogeneous step, the PUs in $F$ have enough memory to get the
desirable block size, but in the remaining three steps, their memory is saturated.
In total, this gives $10$ experiments per graph.
These experiments are performed for $k=24\cdot 2^i$, where $i \in \{2,3\}$.

\begin{table}
\centering
\begin{tabular}{c|c|c|c}
exp & speed & memory & $tw$(fast)$/tw$(slow)\\
\hline
1 & 1 & 2 & 1 - 1 \\
2 & 2 & 3.2 & 2 - 2 \\
3 & 4 & 5.2 & 3.2 - 3.5 \\
4 & 8 & 8.5 & 5.5 - 6.1 \\
5 & 16 & 13.8 & 9.4 - 11.5
\end{tabular}
\caption{Speed and memory of the fast PUs used for \topoA and \topoB.
The slow PUs have computational speed $1$ and memory $2$ for all the experiments.
The last column show approximately how bigger is the target block weight for the fast PUs
for the experiments where $|F|=p/12$ and $p/6$.}
\label{table:topo}
\end{table}

\begin{figure*}[t]
\centering
\begin{subfigure}[b]{.5\textwidth}
  \centering
  \includegraphics[width=8.5cm,height=10cm]{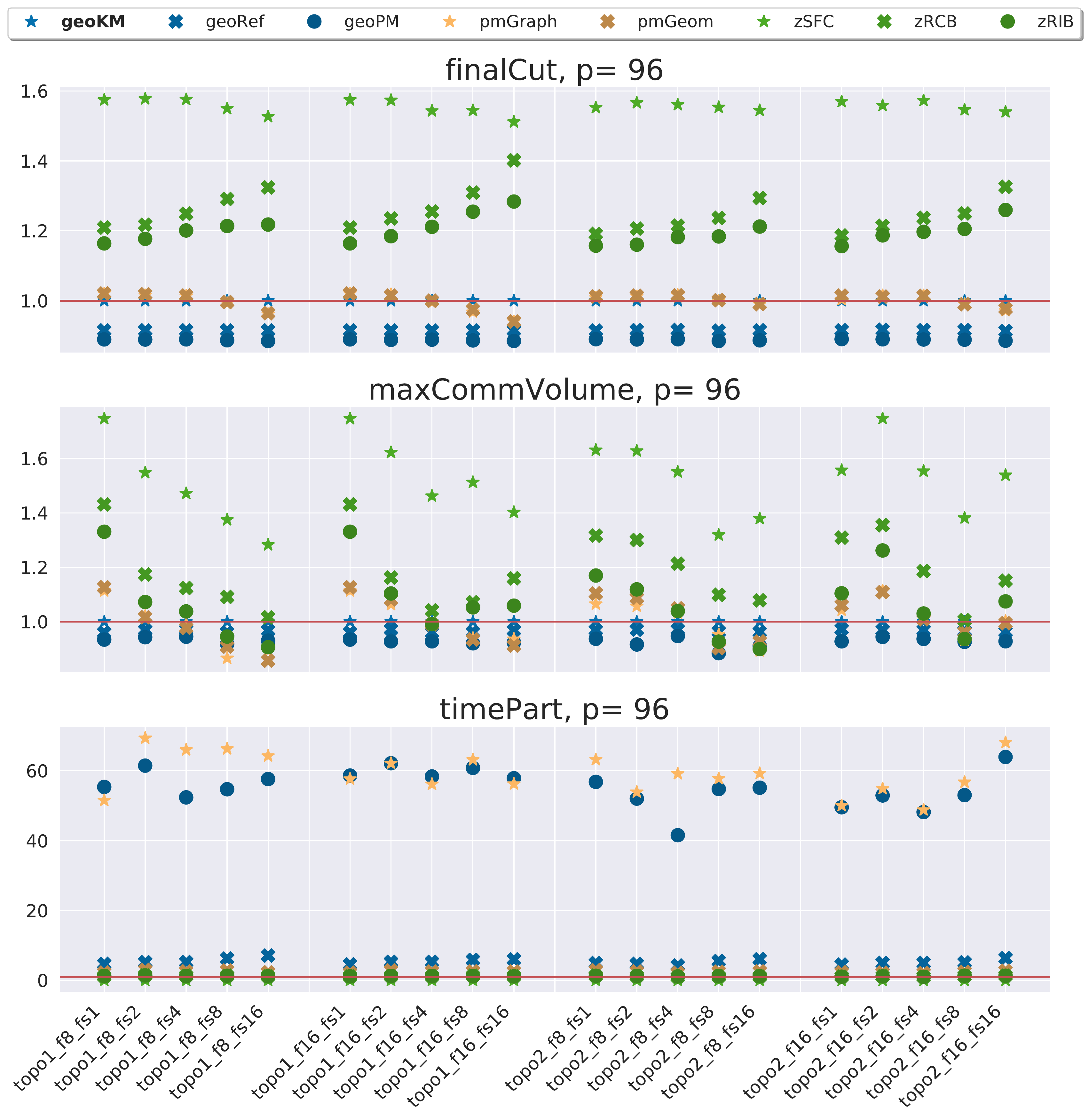}

  \caption{hugeX graphs representing numerical simulation meshes (2D)}
\end{subfigure}%
\begin{subfigure}[b]{.5\textwidth}
  \centering
  \includegraphics[width=8.3cm,height=9.6cm]{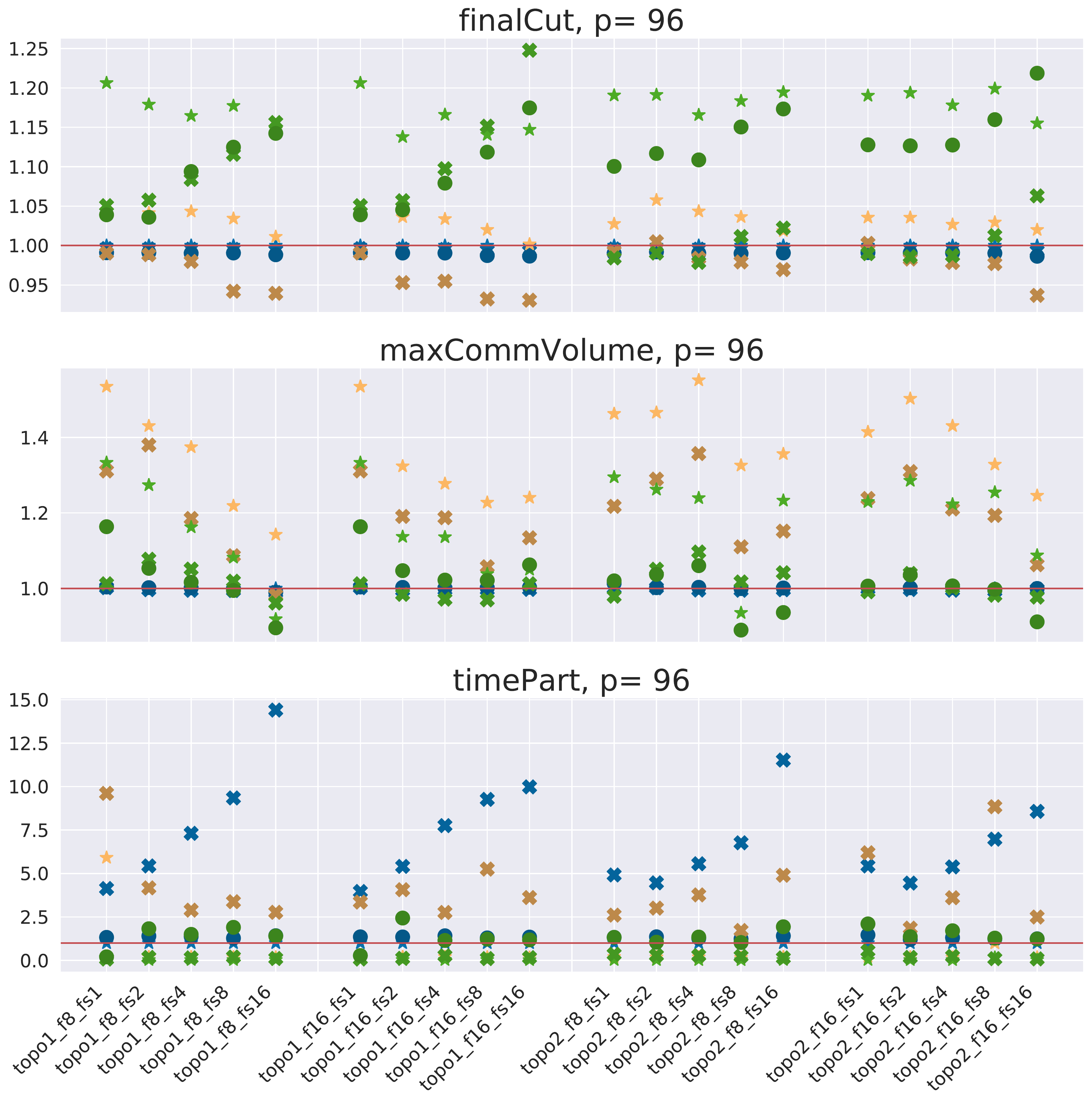}

  \caption{Two \emph{alya} graphs (3D)}
\end{subfigure}%

\caption{Results for (a) three hugeX graphs from the 10th DIMACS Challenge and (b) two alya graphs
using 96 PUs and 16 different topologies.
The data points are the geometric mean of the values for each graph
and are relative to the respective balanced \km~value. 
For the names in the $x$-axis: topoX indicates \topoA or \topoB, `f' indicates the number of
fast PUs and `fs' indicates the speed of the fast PUs. The slow PUs have speed $1$ in all cases.
These experiments were conducted on the local cluster. 
Values are relative to balanced \km\ (lower is better).
}
\label{fig:numSim_huge_alya}
\end{figure*} 

\Cref{fig:numSim_huge_alya}(a) shows results aggregated over the three hugeX graphs.
The $x$-axis shows the different variations of \topoA, the $y$-axis relative quality
or partitioning time.
A first observation is that heterogeneity plays a role, but not a big one. In particular
the geometric methods from Zoltan are negatively affected in terms of quality with increasing heterogeneity
(from left to right).
The geometric-only approaches provide solutions with comparably high cut and max communication volume
values, especially for two dimensions.
Among these, the balanced \km~approach has the best quality in most cases with more than $15\%$ improvement.
Quality-wise, in most cases, \parm~offers solution quality very close to the quality of balanced \km.
But the best solutions come from the two version \geoRef~and \geoPM, which combine 
balanced \km\ with local refinement.

For the 3D \emph{alya} graphs in \Cref{fig:numSim_huge_alya}(b), the
geometric methods narrow the gap somewhat. These application graphs are presented to show results
against the general trend. In a few cases, the geometric methods even give here the solution with the lowest
cut or volume, \eg for \emph{topo1\_f8\_fs16}.
\parm~solutions have good cut values -- on the other hand, the max communication volume is worse,
even compared to geometric methods.
As for \geo, \geoKM~is better than the other geometric methods, but the combination
with refinement offers some improvement.
As expected, geometric methods have the lowest running time, less than a few seconds.
The refinement of \geoRef\ is faster than the refinement of \parm~for the 2D graphs.
For the 3D \emph{alya} graphs, our refinement algorithm is affected by heterogeneity;
the more heterogeneous the system, the slower the algorithm becomes.
This is probably because, as speed and memory of the fast PUs increase,
the size of the blocks that correspond to the fast PUs increases, too.
As we consider a percentage of vertices from each block for the local refinement,
more vertices are eligible to move.

\subsection{Experiments for \topoB}
\label{exp-topoB}

The second category consists of three sets of PUs, $F$, $S_1$ and $S_2$.
It is designed to model systems with two kinds of CPUs and one kind of GPU.
Again, $|F|=k/6$ or $|F|=k/12$, 
but now, the slow PUs are divided in two equally sized groups $|S_1|=|S_2|$.
The PUs in $S_1$ have constant memory and speed, but the speed of the PUs in $S_1$ 
increases as shown in \Cref{eq:speedIncr}:
\begin{equation}
\frac{\cs(s_1)}{\mcap(s_1)} = \frac{1}{2} \frac{\cs(f)}{\mcap(f)}, \;
s_1 \in S_1,\, f\in F
\label{eq:speedIncr}
\end{equation}
This ensures that PUs in $S_1$ will be assigned their block weights after the PUs in $F$ 
but before the PUs in $S_2$.
The specs of the PUs in $F$ are increased as before, see \Cref{table:topo}.
These experiments are performed for $k=24\cdot 2^i$, $i=1,2,\dots,8$.

\begin{figure}[h]
\centering
\includegraphics[scale=0.26]{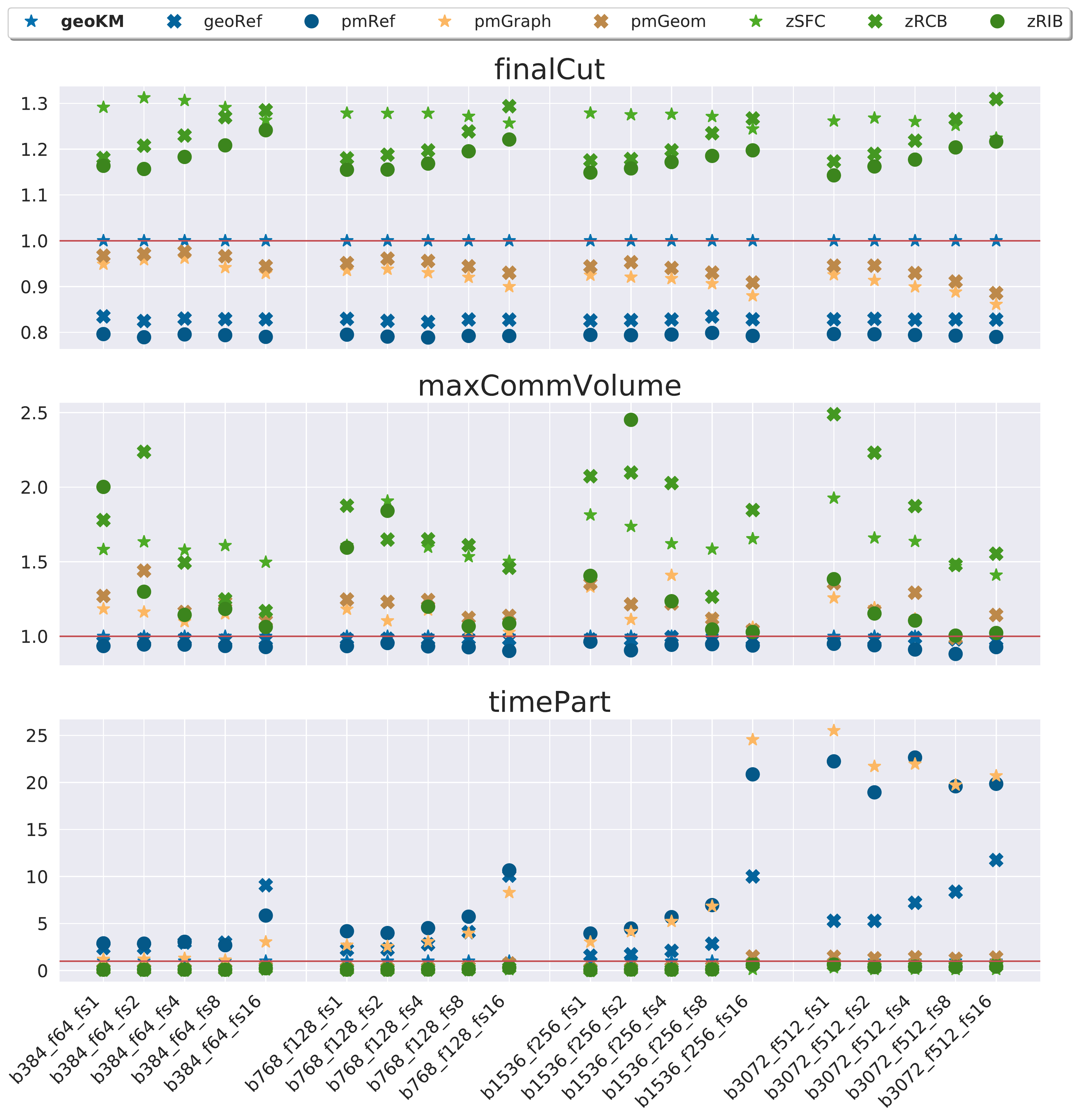}

\caption{Results for the refinetrace graph with $24\cdot 2^i, i=4,\dots,9$ PUs.
All experiments belong to \topoB.
For the names in the x axis: `b' is the number of blocks/PUs, next, `f' indicated the number of
fast PUs and `fs' indicates the speed of the fast PUs. The slow PUs have speed $1$ in all cases.
These experiments were conducted in the Lise cluster.
}
\label{fig:refined_lise}
\end{figure} 

Experiments with high numbers of PUs were conducted on Lise; their results are shown
in \Cref{fig:refined_lise} for the \emph{refinedtrace} graph, which shows representative results:
the algorithms \geoRef\ and \geoPM\ yield
partitions with the lowest cut and communication volume.
As the number of PUs increases, we observe that \parmgr~and \geoPM~suffer from 
high running times, in contrast to our refinement implementation.
Geometric approaches are steadily worse in quality but very fast in running time.
Heterogeneity has again a negative effect on the edge cut produced by the geometric tools from Zoltan,
whereas \parm\ seems to benefit slightly. An effect on the different algorithms in 
\geo\ is hardly visible.

\begin{figure*}[h]
\centering
\includegraphics[scale=0.34]{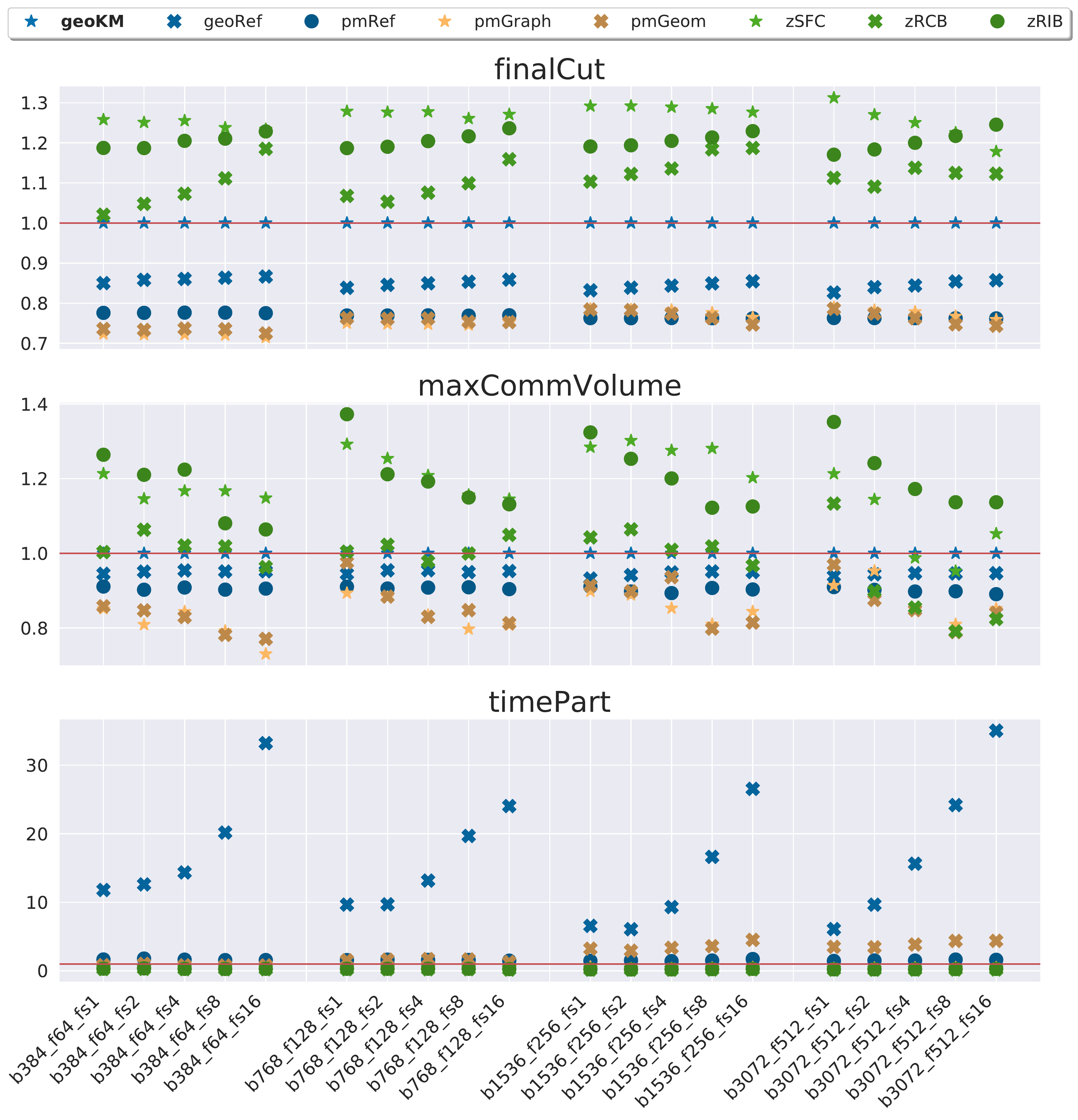}

\caption{Results for the 3D \emph{rgg} and \emph{rdg} graphs with $2^{29}$ vertices,
using $24\cdot 2^i,$ $i=4, \dots, 9$, PUs.
The data points show the geometric mean of the value  for each graph
and are relative to the corresponding balanced \km~value.
All experiments belong to \topoB.
Regarding the names on the $x$ axis: `b' is the number of blocks/PUs, `f' indicates the number of
fast PUs, and `fs' indicates the speed of the fast PUs. The slow PUs have speed $1$ in all cases.
These experiments were run on Lise.
}
\label{fig:r3d_lise}
\end{figure*}	

In \Cref{fig:r3d_lise}, we see experiments with 3D random geometric and Delaunay triangulation graphs.
Regarding quality, the same pattern as before emerges: \geoKM\ is better than other geometric
methods and the combinatorial algorithms all give better solutions (of similar quality).
Again, \geoRef\ seems to suffer from high running times when the topologies become more heterogeneous.

\subsection{Experiments for \topoC}
\label{exp-topoC}
\begin{figure}
\centering
\includegraphics[scale=0.255]{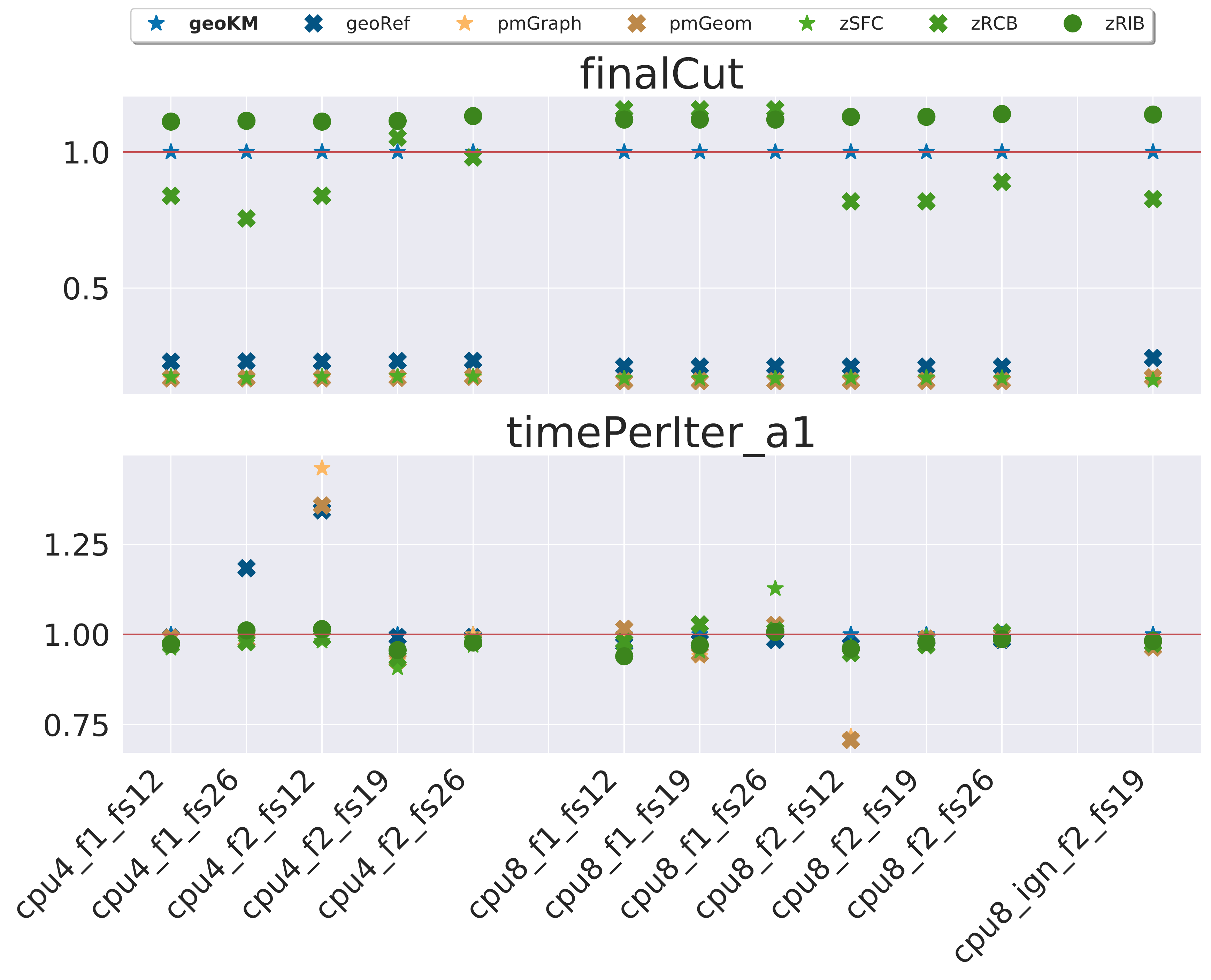}

\caption{Cut values and time per CG iteration using different settings within \topoC
for the \emph{rdg\_2d\_29} graph.
}
\label{fig:rdgCG}
\end{figure}

The third category consists of heterogeneous topologies simulated in the local compute cluster.
We cannot change the specs of an individual core, but we change
all 24 cores of a node. 
The experiments involve 4 and 8 compute nodes (96 and 182 PUs) where, each time, 
$1$ or $2$ nodes (24 or 48 PUs) are unchanged and the rest have their speed and memory lowered in order to
represent the slow PUs. For these experiments we are able to get meaningful
running times for SpMV and CG.
For the benchmarks to be meaningful, we use a sufficiently large graph that still fits into the memory,
rdg\_2d\_29.
In \Cref{fig:rdgCG} we see the edge cut and the running time of the CG solver per iteration for the
linear systems (the SpMV results are similar to CG and thus omitted).
The previous trend regarding the cut quality continues. However,
while the quality in terms of edge cut differs across different tools,
the communication volume and the running time per CG iteration show smaller deviations.
We conjecture that a more severe heterogeneity will also translate into higher real-world differences.

\section{Conclusions}
\label{sec:conclusions}
So far, when working with established partitioning tools for load distribution in sparse matrix / graph problems
on compute systems with different PUs and memory capacities, an additional preprocessing step
is required. This preprocessing computes the different block sizes in a partition, which are
then fed into the partitioners. For the first phase of this two-stage process, we proposed 
a greedy algorithm and proved it to be optimal. Our experiments for the second phase 
indicate that a combination of
geometric and combinatorial methods yield the most promising results regarding the tradeoff between quality
and running time. Overall, the results suggest that more work is necessary to support strongly heterogeneous
computations out of the box and in a scalable way.
This particularly includes a one-phase approach for even more complex scenarios with different
computational weights and communication costs.

\begin{table*}
\centering
\begin{center}
\textbf{} 
\end{center}
\begin{tabular}{cc||cccc|cccc|cccc}
Graphs & competitors & \multicolumn{4}{|c}{finalCut} &  \multicolumn{4}{|c}{maxCommVolume} &  \multicolumn{4}{|c}{timePart} \\ \hline 
 & & t1\_f8 & t1\_f16 & t2\_f8 & t2\_f16 & t1\_f8 & t1\_f16 & t2\_f8 & t2\_f16 & t1\_f8 & t1\_f16 & t2\_f8 & t2\_f16 \\ \hline 
\parbox[t]{2mm}{\multirow{8}{*}{\rotatebox[origin=c]{90}{333SP}}}
 & geoKM & 43428 & 40230 & 41978 & 40108 & 2692 & 1740 & 2395 & 1422 & 1.96 & 2.43 & 2.96 & 28 \\ 
 & geoRef & 42393 & 39294 & 40904 & 39025 & 2623 & 1701 & 2241 & 1342 & 14.12 & 13.26 & 14.87 & 12.76 \\ 
 & geoPM & 41349 & 38591 & \textbf{39967} & \textbf{38365} & 2561 & 1671 & 1822 & \textbf{1321} & 114.13 & 97.47 & 101.5 & 102.77 \\ 
 & pmGraph & \textbf{40291} & 38812 & 44078 & 41311 & \textbf{1839} & 1537 & \textbf{1514} & 1567 & 51.92 & 51.59 & 55.36 & 534 \\ 
 & pmGeom & 40562 & \textbf{38249} & 43109 & 41635 & 1897 & \textbf{1474} & 1554 & 1477 & 3.75 & 10.18 & 8.72 & 8.15 \\ 
 & zSFC & 96465 & 86767 & 99712 & 94397 & 2722 & 3366 & 3486 & 3135 & \textbf{0.04} & \textbf{0.04} & \textbf{0.03} & \textbf{0.03} \\ 
 & zRCB & 90126 & 89530 & 80907 & 75655 & 6373 & 6133 & 4814 & 3236 & 2.46 & 2.85 & 2.46 & 2.77 \\ 
 & zRIB & 52620 & 50356 & 51820 & 52117 & 2474 & 2168 & 1270 & 1556 & 2.48 & 3.49 & 4.94 & 2.83 \\ 
\hline 

\parbox[t]{2mm}{\multirow{8}{*}{\rotatebox[origin=c]{90}{NLR}}}
 & geoKM & 64093 & 61001 & 66647 & 65304 & 2549 & 2110 & 1943 & 2017 & 3.18 & 3.3 & 2.49 & 2.1 \\ 
 & geoRef & 62438 & 59539 & 64986 & 63680 & 2489 & 2066 & 1873 & 1957 & 18.42 & 16 & 16.63 & 12.94 \\ 
 & geoPM & \textbf{61211} & \textbf{58267} & \textbf{63853} & \textbf{62544} & 2443 & 1978 & 1832 & \textbf{1838} & 163.79 & 147.11 & 145.51 & 1483 \\ 
 & pmGraph & 68397 & 63406 & 71938 & 68521 & 2513 & 2160 & 1980 & 2002 & 183.81 & 263.75 & 299.36 & 130.46 \\ 
 & pmGeom & 67995 & 61760 & 71583 & 68238 & 2302 & \textbf{1923} & 2094 & 1886 & 7 & 8.86 & 7.58 & 4.57 \\ 
 & zSFC & 93975 & 89967 & 97693 & 95187 & 3488 & 2686 & 2455 & 2664 & \textbf{0.03} & \textbf{0.03} & \textbf{0.03} & \textbf{0.04} \\ 
 & zRCB & 78400 & 79242 & 78031 & 79429 & \textbf{2061} & 1970 & \textbf{1797} & 1929 & 2.81 & 3.13 & 3.56 & 2.63 \\ 
 & zRIB & 77679 & 77681 & 80737 & 79637 & 2252 & 2059 & 2028 & 1875 & 3.3 & 3 & 3.27 & 2.76 \\ 
\hline 

\parbox[t]{2mm}{\multirow{8}{*}{\rotatebox[origin=c]{90}{hugebubbles-00020}}}
 & geoKM & 55607 & 53167 & 57706 & 55905 & 3567 & 3241 & 2924 & 2673 & 4.73 & 4.22 & 4.43 & 3.67 \\ 
 & geoRef & 46059 & 43728 & 47666 & 46039 & 3459 & 3124 & 2798 & 2609 & 37.52 & 33.37 & 31 & 26.77 \\ 
 & geoPM & \textbf{43796} & \textbf{41536} & \textbf{45379} & \textbf{43754} & 3259 & 2910 & \textbf{2623} & \textbf{2399} & 249.95 & 182.26 & 197.8 & 181.28 \\ 
 & pmGraph & 49102 & 45169 & 51523 & 49960 & 3004 & 2930 & 3165 & 2778 & 319.49 & 1486 & 171.87 & 241.42 \\ 
 & pmGeom & 48998 & 44915 & 51663 & 49568 & \textbf{2832} & \textbf{2758} & 3120 & 2562 & 7.42 & 5.77 & 6.57 & 7.86 \\ 
 & zSFC & 68678 & 62895 & 71609 & 68779 & 4085 & 3675 & 3532 & 3467 & \textbf{0.14} & \textbf{0.14} & \textbf{0.13} & \textbf{0.13} \\ 
 & zRCB & 72584 & 73623 & 78029 & 74534 & 3287 & 3269 & 3283 & 2890 & 3.31 & 3.28 & 3.38 & 3.73 \\ 
 & zRIB & 69821 & 69446 & 71269 & 72760 & 3529 & 3996 & 3026 & 3464 & 3.77 & 3.16 & 3.51 & 3.7 \\ 
\hline 

\parbox[t]{2mm}{\multirow{8}{*}{\rotatebox[origin=c]{90}{hugetrace-00020}}}
 & geoKM & 49064 & 46688 & 51352 & 48714 & 2957 & 2605 & 2509 & 2093 & 2.37 & 2.93 & 2.75 & 2.34 \\ 
 & geoRef & 40330 & 38600 & 42378 & 39911 & 2864 & 2531 & 2447 & 2029 & 26.6 & 19.73 & 20.84 & 209 \\ 
 & geoPM & \textbf{38063} & \textbf{36282} & \textbf{40150} & \textbf{37823} & 2667 & 2359 & 2318 & \textbf{1950} & 153 & 226.38 & 154.79 & 1431 \\ 
 & pmGraph & 41824 & 38749 & 44311 & 42069 & 2831 & 2368 & \textbf{2096} & 2127 & 142.34 & 204.86 & 180.49 & 208.21 \\ 
 & pmGeom & 42079 & 39037 & 45245 & 42701 & \textbf{2450} & \textbf{2310} & 2125 & 2252 & 6.36 & 6.43 & 79 & 99 \\ 
 & zSFC & 56457 & 52976 & 59297 & 56859 & 3574 & 3184 & 3374 & 2711 & \textbf{0.09} & \textbf{0.13} & \textbf{0.09} & \textbf{0.09} \\ 
 & zRCB & 61547 & 62981 & 64210 & 62905 & 3331 & 2804 & 2755 & 2524 & 2.99 & 3.21 & 3.33 & 35 \\ 
 & zRIB & 58929 & 57615 & 61530 & 60848 & 2776 & 2590 & 2500 & 2323 & 2.95 & 37 & 3.34 & 3.33 \\ 
\hline 

\parbox[t]{2mm}{\multirow{8}{*}{\rotatebox[origin=c]{90}{rdg\_2d\_23}}}
 & geoKM & 169940 & 159700 & 177770 & 170440 & 6540 & 5144 & 5777 & 4590 & 0.74 & 0.62 & 0.8 & 0.58 \\ 
 & geoRef & 128020 & 122760 & 132780 & 129850 & 5689 & 4448 & 4899 & 4017 & 10.47 & 6.71 & 7.56 & 5.55 \\ 
 & geoPM & \textbf{102700} & \textbf{97184} & \textbf{107750} & \textbf{104440} & \textbf{4289} & \textbf{3445} & \textbf{3519} & \textbf{3151} & 1.19 & 1.15 & 1.15 & 0.98 \\ 
 & pmGraph & 111690 & 100240 & 115650 & 111950 & 6005 & 3988 & 3763 & 3826 & 0.41 & 0.4 & 0.39 & 0.37 \\ 
 & pmGeom & 110190 & 103660 & 114760 & 112070 & 4331 & 4114 & 3783 & 3453 & 96.25 & 26.48 & 26.6 & 41.46 \\ 
 & zSFC & 121460 & 114430 & 134600 & 128600 & 3270 & 3506 & 3987 & 3488 & \textbf{0.04} & \textbf{0.05} & \textbf{0.04} & \textbf{0.04} \\ 
 & zRCB & 172660 & 184170 & 183960 & 193760 & 5413 & 5181 & 4020 & 4887 & 0.1 & 0.08 & 0.09 & 0.07 \\ 
 & zRIB & 191780 & 181910 & 197910 & 196270 & 5647 & 4810 & 4921 & 4451 & 0.35 & 0.2 & 0.45 & 0.26 \\ 
\hline 

\parbox[t]{2mm}{\multirow{8}{*}{\rotatebox[origin=c]{90}{alyaTestCaseB}}}
 & geoKM & 2066100 & 1949600 & 2171400 & 2081900 & 93325 & 74778 & 72500 & 68538 & 5.13 & 5.64 & 4.97 & 5.59 \\ 
 & geoRef & 2055800 & 1939300 & 2162500 & 2070500 & 92616 & \textbf{74530} & \textbf{72225} & 68407 & 74.41 & 60.75 & 53.32 & 49.3 \\ 
 & geoPM & 2046800 & 1932000 & 2155500 & 2061100 & 92653 & 74847 & 72485 & 68530 & 7.62 & 7.42 & 6.85 & 6.82 \\ 
 & pmGraph & 2085100 & 1956800 & 2234200 & 2140100 & 103480 & 100010 & 98172 & 85985 & 0.9 & 0.87 & 0.87 & 0.88 \\ 
 & pmGeom & \textbf{1955700} & \textbf{1815200} & \textbf{2121700} & \textbf{1949500} & 101460 & 84799 & 78720 & 71147 & 6.85 & 7.27 & 9.98 & 16.22 \\ 
 & zSFC & 2358300 & 2213100 & 2579300 & 2393200 & 82725 & 78263 & 84616 & 72471 & \textbf{0.15} & \textbf{0.16} & \textbf{0.16} & \textbf{0.17} \\ 
 & zRCB & 2382200 & 2429400 & 2185500 & 2190800 & 89201 & 76273 & 74508 & 66581 & 0.6 & 0.61 & 0.74 & 0.48 \\ 
 & zRIB & 2262400 & 2226600 & 2299900 & 2391100 & \textbf{76607} & 76252 & 74954 & \textbf{61899} & 51 & 3.28 & 4.83 & 5.47 \\ 
\hline 

\end{tabular}
\caption{Exact values for a subset of topologies and graphs for 3 metrics: edge cut, maximum communication volume and partitioning time. The results are for 96 PUs and we include 4 topologies, 2 from \topoA and 2 from \topoB with 8 or 16 fast PUs. For the presented results, all fast PUs have a speed of 16 (fs16).
Best results are marked in bold font.}
\label{table:exact_values}\end{table*}

\begin{small}
\paragraph*{Acknowledgments}
This work is partially supported by German Research Foundation (DFG) 
grant ME 3619/4-1. We also acknowledge support by the North-German
Supercomputing Alliance (HLRN).

\end{small}

\bibliography{refs,Meyerhenke}

\iftoggle{arxiv_ver}{

\balance

\clearpage

\appendix

\section{Notation}

\section{Omitted Proofs}
\label{app:omitted-proofs}

\subsection{Proof of Lemma~\ref{lem:change-only-once}}
\label{app:proof-lemma-only-once}

\begin{proof}
We will prove that if some arbitrary PU in position $i$ in the sorted sequence is 
non-saturated, then all PUs in positions greater than $i$ are also non-saturated.
Towards a contradiction, let $d$ be the first iteration in which the first saturated PU
appears \emph{after} a non-saturated one. Consequently, when denoting the state of
\cl\ and \js\ at the beginning of iteration $i$, $0 \leq i \leq k-1$, by $\cl^{(i)}$ and $\js^{(i)}$ respectively, we get:
\begin{align}
\desW(b_d) & > \mcap(p_d) \nonumber \\
\Longrightarrow \cs(p_d) \cdot \frac{\cl^{(d)}}{\js^{(d)}} & > \mcap(p_d) \nonumber \\
\Longrightarrow \frac{\cs(p_d)}{\mcap(p_d)} & > \frac{\js^{(d)}}{\cl^{(d)}} \label{eq:saturated-fraction}
\end{align}

By construction PU $p_{d-1}$ is not saturated. Thus:
\begin{align}
\desW(b_{d-1}) & \leq \mcap(p_{d-1}) \nonumber  \\
\Longrightarrow \frac{\cs(p_{d-1})}{\mcap(p_{d-1})} & \leq \frac{\js^{(d-1)}}{\cl^{(d-1)}} \label{eq:no-satur}
\end{align}

Since the PUs are sorted non-increasingly (and with \Cref{eq:saturated-fraction}), we obtain:
\begin{align} 
\frac{\cs(p_{d-1})}{\mcap(p_{d-1})} \geq \frac{\cs(p_d)}{\mcap(p_d)} 
> \frac{\js^{(d)}}{\cl^{(d)}} \label{eq:contrad1} 
\end{align}

By Line~\ref{line:subtract-block-size}, the computational load left at the end
of iteration $d$
is what was left beforehand minus what we assigned to PU $p_d$ -- similarly for the joint speed:
\begin{align*}
\cl^{(d)} = \cl^{(d-1)} - \tw(b_{d-1}) \\
\js^{(d)} = \js^{(d-1)} - \cs(p_{d-1}) 
\end{align*}

Let us take a closer look to $\cl^{(d)}$. Since PU $p_{d-1}$ is non-saturated,
$\tw(p_{d-1}) = \desW(b_{d-1})$, so that we get:
\begin{align}
\cl^{(d)} = & \cl^{(d-1)} - \tw(b_{d-1}) \nonumber \\
 = & \cl^{(d-1)} - \cs(p_{d-1}) \cdot \frac{\cl^{(d-1)}}{\js^{(d-1)}} \nonumber \\
 = & \cl^{(d-1)} \cdot \frac{\js^{(d-1)} - \cs(p_{d-1})}{\js^{(d-1)}} \nonumber \\
 = & \cl^{(d-1)} \frac{\js^{(d)}}{\js^{(d-1)}} \label{eq:contrad2}
\end{align}

Combining \Cref{eq:contrad1} and \Cref{eq:contrad2} yields:
\begin{align}
& \frac{\cs(p_{d-1})}{\mcap(p_{d-1})} > \frac{\js^{(d)}}{\cl^{(d)}} 
= \frac{\js^{(d)}}{ \cl^{(d-1)} \cdot \frac{\js^{(d)}}{\js^{(d-1)}} }  \nonumber \\
= & \frac{\js^{(d)} \js^{(d-1)}}{ \cl^{(d-1)} \js^{(d)} } = \frac{ \js^{(d-1)}}{ \cl^{(d-1)}},
\end{align}
which is the desired contradiction of \Cref{eq:no-satur}.
\end{proof}

\subsection{Proof of Theorem~\ref{thm:optimal-load}}
\label{app:proof-opt-thm}
\begin{proof}
The running time is dominated by the sorting step, since the loop has $k$ iterations and all other
operations need only constant time.

Clearly, all saturated PUs receive their optimal load, since more load would exceed their memory
capacity. Let $d$ now be the iteration in which the first non-saturated PU appears in our sorted
sequence. We need to show that all PUs $p_i$ with $i \geq d$ receive a load that is proportionate to
$\cl^{(d)}  \cdot \frac{\cs(p_d)}{\js^{(d)}}$ -- as this minimizes Eq.~(\ref{eq:min-imb}).

Let us apply an inductive argument, starting in iteration $d$. 
Clearly, PU $\tw(b_d)$ fulfills this condition. For the inductive step, we assume the claim
to be true for all iterations $i'$ with $d \leq i' \leq i$. We now consider iteration $i+1$:
\begin{align*}
tw(b_{i+1}) & = \cs(p_{i+1}) \cdot \frac{\cl^{(i)} - tw(b_i)}{\js^{(i)} - \cs(p_i)} \\
\end{align*}
In order to prove that the assignment in iteration $i+1$ is proportionate to the load and speed situation
in iteration $i$ (and thus $d$ by our assumption), it is sufficient to show that
\[
\frac{\cl^{(i)} - tw(b_i)}{\js^{(i)} - \cs(p_i)} = \frac{\cl^{(i)}}{\js^{(i)}}.
\]
Inserting the value for $tw(b_i)$ and some rearranging yields:
\begin{align*}
\label{eq:sum-of-loads2}
& \frac{\cl^{(i)} - tw(b_i)}{\js^{(i)} - \cs(p_i)} \\
= & \frac{\cl^{(i)} - \left(\frac{\cl^{(i)} \cdot \cs(p_i)}{\js^{(i)}}\right)}{\js^{(i)} - \cs(p_i)} \\
= & \frac{\js^{(i)} \left( \cl^{(i)} - \left(\frac{\cl^{(i)} \cdot \cs(p_i)}{\js^{(i)}}\right) \right)}{\js^{(i)} \left( \js^{(i)} - \cs(p_i) \right)} \\
= & \frac{\cl^{(i)}}{\js^{(i)}} \left( \frac{\js^{(i)} - \cs(p_i)}{\js^{(i)} - \cs(p_i)} \right) \\
= & \frac{\cl^{(i)}}{\js^{(i)}}.
\end{align*}
\end{proof}

}

\end{document}